\documentclass{JFM-FLM_Au}

\usepackage[LGR,T1]{fontenc}
\usepackage[utf8]{inputenc}
\usepackage{textalpha}

\usepackage{amssymb}
\usepackage{graphicx}
\usepackage{amsmath} 
\usepackage{subcaption} 
\usepackage{overpic} 
\usepackage{bm}   
\usepackage{gensymb} 
\usepackage[separate-uncertainty = true]{siunitx} 
\usepackage{tikz} 
\usepackage{xr} 
\usepackage{hyperref} 
\usepackage{cleveref} 
\usepackage{adjustbox}  
\usepackage[inline]{enumitem}

\graphicspath{{Figures/}}

\title{Fractal forcing of turbulence by settling particles}

\lefttitle{S. Tandurella, A. Chiarini and M.E. Rosti}
\righttitle{Journal of Fluid Mechanics}

\author{Simone Tandurella\aff{1}, Alessandro Chiarini\aff{2} \and Marco Edoardo Rosti\aff{1}}

\affiliation{\aff{1}Complex Fluids and Flows Unit, Okinawa Institute of Science and Technology Graduate University, 1919-1 Tancha, Onna-son, Okinawa 904-0495, Japan
\aff{2}Dipartimento di Scienze e Tecnologie Aerospaziali,	Politecnico di Milano,  via La Masa 34, 20156 Milano, Italy}

\corresau{Marco Edoardo Rosti, \email{marco.rosti@oist.jp}}

\begin{document}

\maketitle
\newcommand{\ddp}{d}
\newcommand{\mdp}{m}
\newcommand{\ndp}{n}
\newcommand{\Vdp}{V}
\newcommand{\Idp}{I}
\newcommand{\rhop}{\rho_{p}}
\newcommand{\rhof}{\rho_{f}}
\newcommand{\Gal}{\textit{Ga}}
\newcommand{\Sto}{\textit{St}}
\newcommand{\rad}{\perp}
\newcommand{\ver}{\parallel}
\newcommand{\urad}{\bm{u}_\rad}
\newcommand{\uver}{\bm{u}_\ver}
\newcommand{\vrad}{v_\rad}
\newcommand{\vver}{v_\ver}
\newcommand{\rrad}{r_\rad}
\newcommand{\rver}{r_\ver}
\newcommand{\krad}{\kappa_\rad}
\newcommand{\kver}{\kappa_\ver}
\newcommand{\Rel}{\Rey_\lambda}
\newcommand{\Ret}{\Rey_t}
\newcommand{\Rep}{\Rey_p}
\newcommand{\rhoratio}{\rho}
\newcommand{\PhiV}{\phi}
\newcommand{\taugrav}{\tau}
\newcommand{\ut}{v_t}
\newcommand{\taug}{\tau_g}
\newcommand{\taut}{\tau_t}
\newcommand{\vg}{\ug}
\newcommand{\vt}{\ut}
\newcommand{\utens}{\mathcal{U}}
\newcommand{\PhiVsm}{\widehat{\Phi}_V}
\newcommand{\plane}{\square}
\newcommand{\Zavg}{{xy}}
\newcommand{\shell}{\odot}
\newcommand{\pars}{\therefore}
\newcommand{\ftop}{{fp}}
\newcommand{\ptof}{{pf}}
\newcommand{\ptop}{{pp}}
\newcommand{\gridsp}{\Delta x}
\newcommand{\PhiZavg}{\langle\PhiV\rangle_\Zavg}
\newcommand{\voro}{{v}}
\newcommand{\cluster}{{c}}

\definecolor{cA1}{HTML}{528BFF}
\definecolor{cB1}{HTML}{40D8F7}
\definecolor{cC1}{HTML}{FFC252}
\definecolor{cD1}{HTML}{F77540}
\newcommand{\mycontline}{\raisebox{2pt}{\tikz{\draw[black,line width=0.9pt](0,0) -- (5.5mm,0);}}}
\newcommand{\mydashedline}{\raisebox{2pt}{\tikz{\draw[black,dashed,line width=0.9pt](0,0) -- (5.5mm,0);}}}
\newcommand{\mydottedline}{\raisebox{2pt}{\tikz{\draw[black,dotted,line width=1.0pt](0,0) -- (4.8mm,0);}}}
\newcommand{\dotA}{\raisebox{-0.7pt}{\tikz{\draw[cA1,fill=cA1] (0,0) circle (0.75ex);}}}
\newcommand{\dotB}{\raisebox{-0.7pt}{\tikz{\draw[cB1,fill=cB1] (0,0) circle (0.75ex);}}}
\newcommand{\dotC}{\raisebox{-0.7pt}{\tikz{\draw[cC1,fill=cC1] (0,0) circle (0.75ex);}}}
\newcommand{\dotD}{\raisebox{-0.7pt}{\tikz{\draw[cD1,fill=cD1] (0,0) circle (0.75ex);}}}

\vspace{-20pt}
\begin{abstract}
{\small The settling of particles in a fluid medium can trigger complex collective dynamics, dramatically affecting the distribution of energy across flow scales.
In this work we present the first fully-resolved observations of the turbulence induced by large particle structures, also known as cluster-induced-turbulence (CIT), obtained through particle-resolved direct numerical simulations at high volume fractions and density ratios in large domains.
We find that particles arrange in macroscopic fractal structures, exciting turbulent fluctuations at scales much larger than the individual constituents. By virtue of their fractal nature, these clusters force the continuous-phase in a multiscale, self-similar way. 
We further show that these cluster-driven, vertically decorrelated fluctuations at large scales coexist with, but are fundamentally distinct from, the established particle-driven high-wavenumber scaling. 
Beyond unveiling a different form of clustering in particle-settling, these results highlight a naturally arising, self-assembling path to fractal-generated turbulence.}
\end{abstract}

\begin{keywords}
\vspace{-15pt}
\end{keywords}

\vspace{-15pt}
\section{Introduction}
The sedimentation of macroscopic particles through a viscous fluid is a canonical problem in fluid mechanics, fundamental to an array of geophysical phenomena \citep{meiburg-kneller2010,sutherland-barrett-ea2015,brandt-coletti2022,marchioli-rosti-ea2026} and industrial processes \citep{kunii-levenspiel1991,fullmer-hrenya2017,hardy-rauchenzauner-ea2024}. Despite its long history, predicting the collective dynamics of settling suspensions remains a formidable challenge. The difficulty stems directly from the highly non-linear, chaotic coupling between the continuous fluid motion and the collective hydrodynamics of the dispersed phase \citep{brandt-coletti2022,peng-karzhaubayev-ea2025}.
A striking manifestation of this complex coupling is the clustering instability \citep{fullmer-hrenya2017}, wherein particles spontaneously self-segregate into dynamically evolving, multiscale structures. While simple dissipative systems naturally form clusters \citep{goldhirsch-zanetti1993}, realistic flows feature a delicate interplay of stabilization and destabilization mechanisms that generate exceptionally rich phenomenology \citep{kunii-levenspiel1991,fullmer-hrenya2017}. By acting as localized sources of momentum, these particle clusters transfer their chaotic dynamics to the continuous. The emerging large-scale excitation of the flow by settling particles is known as cluster-induced turbulence \citep[CIT,][]{capecelatro-desjardins-ea2015}.

To date, however, CIT has remained largely inaccessible at a fundamental level. Experimentally, spatially resolved, interference-free measurements of the interstitial fluid within dense suspensions are notoriously difficult. Numerically, CIT has been explored only through coarse-grained, closure-based models, due to the computational cost of resolving fluid-solid interfaces \citep{capecelatro-desjardins-ea2015}. Conversely, fully resolved numerical investigations of spontaneous clustering have been mostly limited to restricted domains or limited portions of the parameter space, yielding strongly correlated structures rather than fully developed CIT \citep{uhlmann-doychev2014,moriche-hettmann-ea2023,jiang-brandt-ea2025,chiarini-gallorini-ea2026,tandurella-rosti-ea2026}. As a result, the fundamental qualities of CIT and the mechanisms driving it persist as open questions.
To address them, we provide a first fully resolved, direct observation of CIT.
We demonstrate that in this regime vertical flow correlation is broken, and genuine CIT characterized by fluid energy excitations at scales vastly exceeding the particle diameter is triggered. 
Through a topological characterization, we uncover that this turbulence is driven by a hierarchy of structures dominated by a single, inherently fractal, domain-spanning cluster. A decomposition of the momentum forcing explicitly shows that this fractal cluster is responsible for nearly all large-scale energy injection into the flow, resulting in a self-similar large-scale fluid forcing.

\vspace{-20pt}
\section{Methods}
\label{sec:methods}
We perform direct numerical simulations of the incompressible Navier-Stokes equations. The fluid velocity field $\bm{u}$ is forced by finite-size particles via a momentum source term $\bm{f}_p=\sum_i^\ndp \bm{F}_p^i$, where $\ndp$ is the total number of particles and $\bm{F}_{p}$ is the interphase coupling force distributed over the particle surface $S$, computed via the immersed boundary method introduced by \cite{hori-rosti-ea2022}. A macroscopic hydrostatic pressure gradient enforces a zero-net-flux condition across the domain. No further external forcings act on $\bm{u}$.
The monodisperse spherical particles evolve according to the Newton-Euler equations of motion. They are driven by the integrated coupling forces $\int_S \bm{F}_p dS$, a constant buoyancy force $(\rhop-\rhof)\Vdp\bm{g}$, and inter-particle collisions handled via a soft-sphere model. Here, $\rhop$ and $\rhof$ are the solid and fluid mass densities, $\bm{g}$ is the vertically oriented gravitational acceleration, and $\Vdp=\pi \ddp^3/6$ is the particle volume, with $\ddp$ being the diameter.
To allow the unconstrained development of mesoscale particle structures---a critical requirement for observing CIT---we utilize a large, vertically elongated triperiodic domain of total volume $V_D=80\ddp\times80\ddp\times200\ddp$. We investigate two primary cases at a high volume fraction $\phi=\ndp\Vdp/V_D=0.10$ ($\ndp=244\,462$) with density ratios $\rho=\rhop/\rhof$ equal to $10$ and $100$, a region of the parameter space where hydrodynamic and collisional effects coexist. Two additional dilute cases ($\phi=0.01$) are discussed in \S\ref{app:phi1}. 
The Galileo number $\Gal=\sqrt{(\rhoratio-1)\ddp^3 |\bm{g}|}/\nu$ is set to $61$ across all cases, corresponding to a single-particle terminal Reynolds number of $\Ret=\vt\ddp/\nu\approx58$ according to empirical correlations \citep{clift-grace-ea2005,tandurella-rosti-ea2026}. Here, $\vt$ is the single-particle terminal velocity and $\nu$ is the kinematic viscosity of the fluid.
The governing equations are discretised and advanced in time at a spatial resolution of $\ddp/\Delta x= 12$, where $\Delta x$ is the grid size; numerical details of the software are reported by \cite{rosti2026}. Comparisons of $\Ret$ as measured from single-particle settling simulations at the same resolution obtain a relative error of $7.6\%$ and $4.8\%$ for the $\rhoratio=10$ and $100$ cases, respectively. Particles are initialized with a uniform random distribution and settle freely until a statistically steady state is reached; all reported statistics are averaged over this post-transient ergodic regime, for an interval of around $600\ddp/\vt$. All non-scalar quantities are decomposed along the gravity-parallel $\ver$ (vertical) and gravity-normal $\rad$ (horizontal) directions: for any vector $\bm{a}=\left(a_x,a_y,a_z\right)$, $\bm{a} = \bm{a}_\rad + \bm{a}_\ver$, where $\bm{a}_\rad=\left(a_x,a_y,0\right)$ and $\bm{a}_\ver=\left(0,0,a_z\right)$. $a_\rad$ and $a_\ver$ indicate $|\bm{a}_\rad|$ and $a_z$, respectively.

\vspace{-5pt}
\section{Results}

\subsection{Flow visualisation and energy spectra}
\begin{figure*}[t!]
  \centering
  \hrule
  \vspace{4pt}
  \begin{subcaptiongroup}
    \begin{overpic}[width=\linewidth]{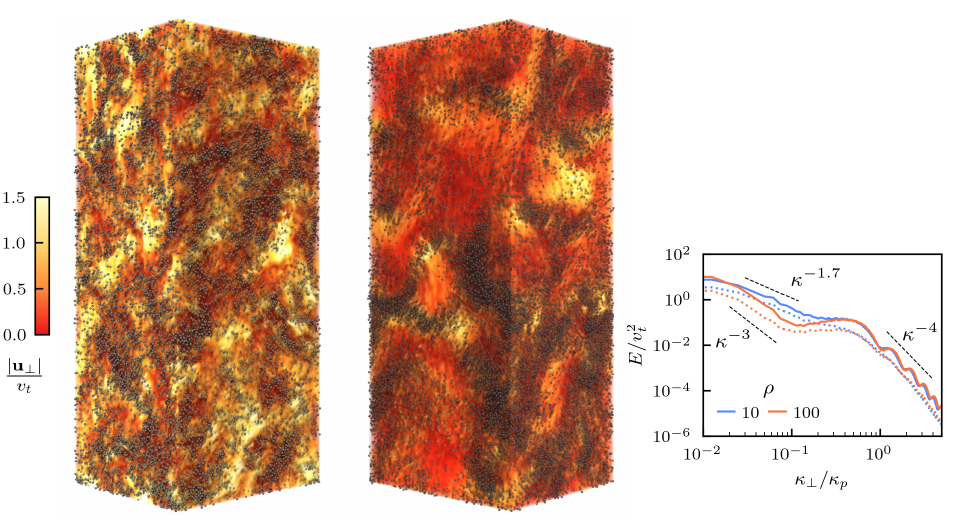}
      \subcaptionlistentry{}
      \label{fig:overview}
      \put(02,53){\captiontext*{}}
      \put(07,52){\tikz\draw[cA1,fill=cA1] (0,0) circle (0.75ex);}
      \subcaptionlistentry{}
      \label{tab:turbulence}
      \put(67,50){\captiontext*{}}
      \put(37,52){\tikz\draw[cD1,fill=cD1] (0,0) circle (0.75ex);}
      \subcaptionlistentry{}
      \label{fig:ekx}
      \put(67,32){\captiontext*{}}
      \put(67,42){
        \begin{adjustbox}{scale=.6} 
        \begin{tabular}{rrccc}
        \specialrule{0.8pt}{2pt}{3pt}
               ID
            &  $\rhoratio$ 
            &  $\frac{\langle K_{\rad}\rangle}{\langle K_{\ver}\rangle}$
            &  ${\langle \Rel\rangle}$             
            &  ${\langle \Sto\rangle}$             \\
        \specialrule{0.5pt}{3pt}{3pt}
        \tikz\draw[cA1,fill=cA1] (0,0) circle (0.75ex);\ \ &$10$  &  $0.54 \pm0.05$  & $33\pm4$& $59.4 \pm 0.7$  \\ 
        \tikz\draw[cD1,fill=cD1] (0,0) circle (0.75ex);\ \ &$100$ &  $0.33 \pm0.06$  & $25\pm2$& $603  \pm13  $ \\ 
        \specialrule{0.8pt}{1pt}{2pt}
        \end{tabular}
        \end{adjustbox}
      }
    \end{overpic}
    \vspace{-12pt}
   \end{subcaptiongroup}
  \hrule
  \captionsetup{subrefformat=parens}
  \caption{Fluid velocity fluctuations. \subref{fig:overview}~Volumetric renderings of the horizontal velocity $|\urad|/\vt$ for $\rho=10$ (left, \protect\dotA\,) and $100$ (right, \protect\dotD\,). \subref{tab:turbulence}~The total kinetic energy isotropy ratio, where $K=\int E(\bm{\kappa})d\bm{\kappa}$, the Taylor microscale Reynolds number $\Rel= {\lambda\sqrt{2K/3}}/{\nu}$, where $\lambda=\sqrt{10\nu K/\varepsilon}$ and $\varepsilon$ is the fluid dissipation rate \citep{pope2000}, and the Stokes number $\Sto=\tau_p/\tau_\eta$, with $\tau_p = \rhoratio\ddp^2/(18\nu)$, and $\tau_\eta = \left(\nu/\varepsilon\right)^{1/2}$. 
  \subref{fig:ekx}~Vertical ($E_\ver$, {\protect\mycontline}) and horizontal ($E_\rad$, {\protect\mydottedline}) energy spectra in the horizontal plane. Reported scalings are approximate visual fits.}
  \label{fig:fig1}
\end{figure*}

\begin{figure*}[ht]
  \centering
  \hrule
  \vspace{1pt}
  \begin{subcaptiongroup}
    \begin{overpic}[width=\linewidth]{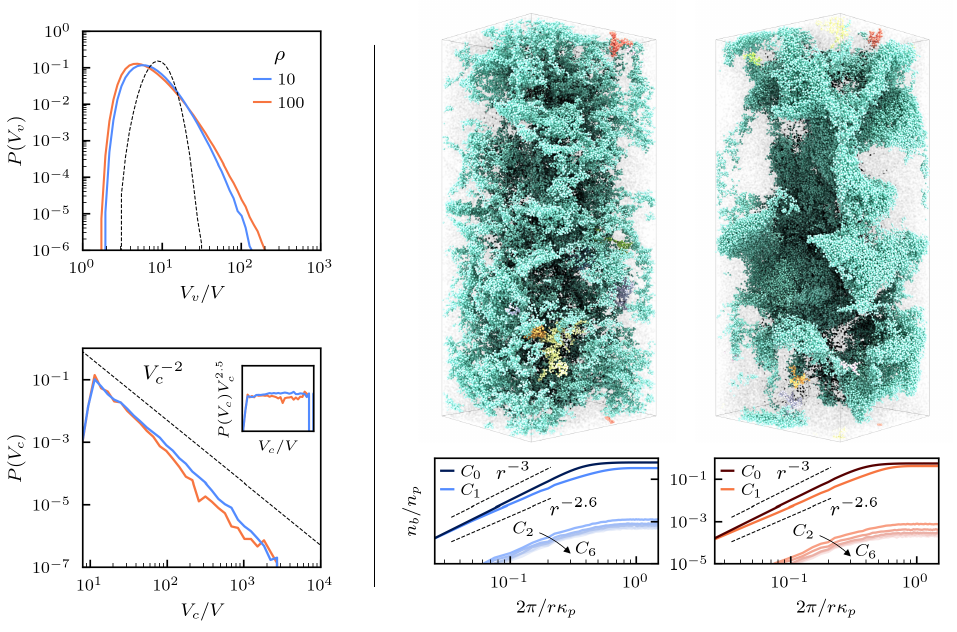}
      \subcaptionlistentry{}
      \label{fig:voronoi}
      \put(00,63){\captiontext*{}}

      \subcaptionlistentry{}
      \label{fig:voroclust}
      \put(00,30){\captiontext*{}}

      \subcaptionlistentry{}
      \label{fig:clusters}
      \put(44.3,63){\tikz\draw[cA1,fill=cA1] (0,0) circle (0.75ex);}
      \put(73,63)  {\tikz\draw[cD1,fill=cD1] (0,0) circle (0.75ex);}
      \put(39,63)  {\captiontext*{}}

      \subcaptionlistentry{}
      \label{fig:fdim}
      \put(39,20){\captiontext*{}}
    \end{overpic}
  \end{subcaptiongroup}
  \hrule
  \captionsetup{subrefformat=parens}
  \caption{Cluster analysis: \subref{fig:voronoi}~Distribution of Voronoi volumes $V_v$. The Poisson reference distribution is shown with a dashed ({\protect\mydashedline}) line. \subref{fig:voroclust}~Distribution of cluster volumes $V_c$. The inset shows the compensated distribution. \subref{fig:clusters}~Cluster visualization for $\rho=10$ (left, \protect\dotA\,) and $100$ (right, \protect\dotD\,). Particles belonging to clusters $C_1$--$C_6$ are colored according to the cluster they belong to, whereas particles that do not belong to any cluster ($C_0$) are shown in semi-transparency. On average $C_1$ accounts for $0.33\phi$ and $0.42\phi$ for $\rhoratio=10$, and $100$, respectively. \subref{fig:fdim}~Number of boxes $n_b$ of size $r$ required to cover each cluster $C_i$.}
  \label{fig:cluster}
\end{figure*}
We begin by examining the instantaneous spatial distribution of the dispersed phase and the resulting fluid velocity field in \cref{fig:overview}. The particles distribute highly heterogeneously, spontaneously organizing into macroscopic clusters. Notably, unlike the vertically coherent, domain-spanning ``streamers'' typically observed in smaller domains or dilute regimes \citep{uhlmann-doychev2014,moriche-hettmann-ea2023,jiang-brandt-ea2025}, these dense clusters exhibit a clear break in vertical symmetry and coherency. This complex spatial arrangement forces the fluid in chaotic patterns. The resulting flow structures extend across a vast range of scales, with macroscopic fluid excitations extending well beyond the size of the constituent particles.

To rigorously quantify the existence of these multiscale fluid motions, we compute the scale-by-scale fluid kinetic energy spectra. Following the system's symmetries, we first define $\bm{E}(\bm{\kappa})=(\hat{u}_x\hat{u}_x^{*},\hat{u}_y\hat{u}_y^{*},\hat{u}_z\hat{u}_z^{*})$, where $\hat{\cdot}=\mathcal{F}(\cdot)$ represents the Fourier transform in the wavenumber vector $\bm{\kappa}$, and $\hat{\cdot}^{*}$ its complex conjugate. We then compute the horizontal variation under planar isotropy assumption, i.e. 
$\bm{E}\left(\krad\right) = \int
\bm{E}\left(\bm{\kappa}\right) \delta_{\rad} d\bm{\kappa}$, where $\delta_{\rad} \equiv \delta(|\bm{\kappa}_\rad|-\krad)$.

$E_\ver(\krad)$ and $E_\rad(\krad)$ are shown in \cref{fig:ekx}, and firmly substantiate our visual observations: while $E_\ver$ dominates the total energy content, $E_\rad$ exhibits a qualitatively similar spectral footprint, confirming the existence of a fully three-dimensional, albeit anisotropic, macroscopic turbulent field (selected turbulence quantities are reported in tab.~\ref{tab:turbulence}).
Most remarkably, however, \cref{fig:ekx} reveals a broadband energy distribution, and two distinct dynamical regimes separated by the particle scale $\kappa_p \equiv 2\pi/\ddp$.
At high wavenumbers ($\krad \gtrsim \kappa_p$), the spectra collapse onto the unique scaling $\kappa^{-4}$ characteristic of particle scale fluctuations \citep[see for example][]{wang-ayala-ea2014,pandey-mitra-ea2023,jiang-brandt-ea2025,tandurella-musacchio-ea2026,fossa-rosti2026}, largely independent of the density ratio. In contrast, at large scales ($\krad < \kappa_p$) we measure a density-dependent accumulation of kinetic energy. The large scale separation from $\kappa_p$ necessarily traces the origin of this accretion to larger structures, i.e. clusters of particles.  

\subsection{Cluster multiscale topology}
To further elucidate the physical origin of these macroscopic fluid fluctuations, we mathematically isolate the particle clusters and analyse their multiscale topology. We start by quantifying the spatial inhomogeneity using a three-dimensional Voronoi tessellation \citep{monchaux-bourgoin-ea2010}. By comparing the probability distribution of the particle Voronoi volumes ($V_\voro$) against a uniformly random distribution of non-overlapping hard spheres at the same volume fraction, we obtain a robust measure of clustering (\cref{fig:voronoi}). Both cases exhibit heavy clustering, marked by strongly skewed distributions where the degree of agglomeration increases with the density ratio $\rho$.
Following standard practice, we define physically coherent clusters as interconnected, vertex-sharing particle regions whose $V_\voro$ falls below the first crossover point with the reference distribution \citep{monchaux-bourgoin-ea2010}. The probability distribution of the resulting macroscopic cluster volumes, $V_\cluster$, is remarkably revealing (\cref{fig:voroclust}). In conditions of dilute, externally forced turbulence, cluster volumes typically follow a power-law decay of $\sim V_\cluster^{-2}$ \citep{monchaux-bourgoin-ea2010,obligado-teitelbaum-ea2014,baker-frankel-ea2017,hassaini-petersen-ea2023}, while inertial-range theoretical arguments predict $\sim V_\cluster^{-16/9}$ \citep{goto-vassilicos2006,yoshimoto-goto2007,fossa-rosti2026}. Under our conditions of dense, spontaneous sedimentation without external forcing, the distribution retains a robust power-law scaling, albeit with a steeper decay $\sim V_\cluster^{-2.5}$ (see inset of \cref{fig:voroclust}), confirming a self-similar, fractal geometry of the clusters. This scaling exponent is invariant across both density ratios, suggesting a distinct geometrical signature of dense, gravity-driven clustering.

Building upon this topological scaling, we now probe the internal geometric structure of the clusters to understand their role in generating fluid fluctuations. Representative spatial reconstructions (\cref{fig:clusters}) reveal a clear structural hierarchy: instead of a uniform dispersion of similarly sized aggregates, the particle phase is dominated by a single, highly connected, domain-spanning primary cluster. The remaining particles form smaller satellite structures interspersed within this dominant network. 
To quantify this hierarchy and confirm the multiscale nature of clusters, we measure their fractal dimension $D_f$ \citep{falconer2013} across statistically independent snapshots (\cref{fig:fdim}). For each realization, clusters are ranked by particle count, with $C_1$ denoting the largest macroscopic cluster, and $C_{i>1}$ representing the progressively smaller ones. The unclustered background particles are categorized as $C_0$. As expected for a homogeneous distribution, the unclustered phase $C_0$ exhibits a clean, three-dimensional scaling ($D_f = 3$). Instead, the primary $C_1$ structure manifests a robust range of self-similar scaling characterized by a strictly lower fractal dimension ($D_f \approx 2.6$). 
Once again, this fractal dimension remains invariant across both density ratios. 
The smaller satellite clusters ($C_{i>1}$) exhibit similar scaling tendencies, though constrained by their limited spatial extent. The identification of the single, domain-spanning, fractal structure ($C_1$) naturally bridges the topological observations with the resulting fluid dynamics, suggesting this structure as the physical source of the multiscale momentum forcing that drives the large-scale turbulence.

\subsection{Multiscale forcing mechanism}
To establish the link between the fractal cluster topology and the large-scale fluid velocity fluctuations, we turn our attention to the particle-mediated conversion of gravitational potential to kinetic energy. To do this, we evaluate the scale-by-scale vertical power flux attributed to the fluid-solid coupling in the kinetic energy budget:  
\par
\begin{figure*}[t]
  \centering
  \begin{subcaptiongroup}
    \begin{overpic}[width=.48\linewidth]{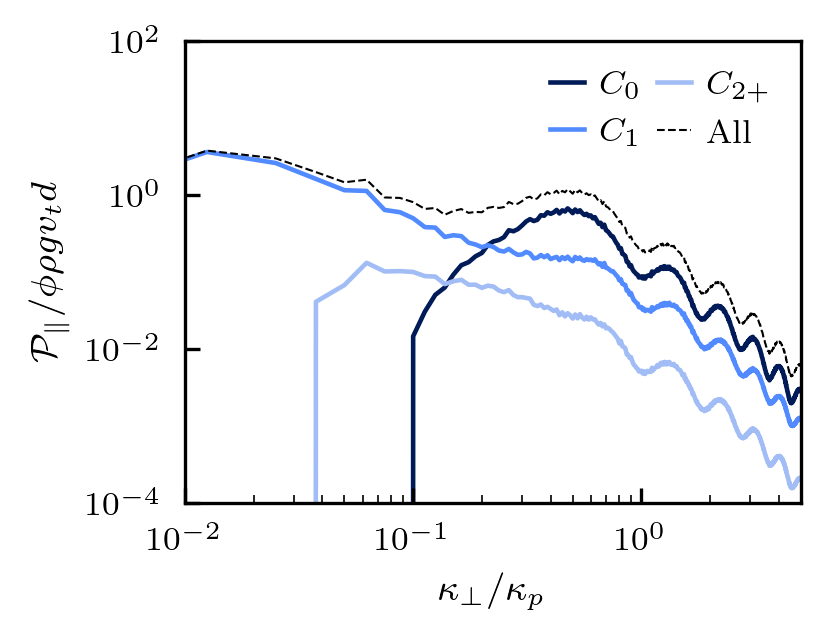}
        \subcaptionlistentry{}
        \put(03,73){\captiontext*{}}
    \end{overpic}%
    \label{fig:cfor10}
    \hfill
    \begin{overpic}[width=.48\linewidth]{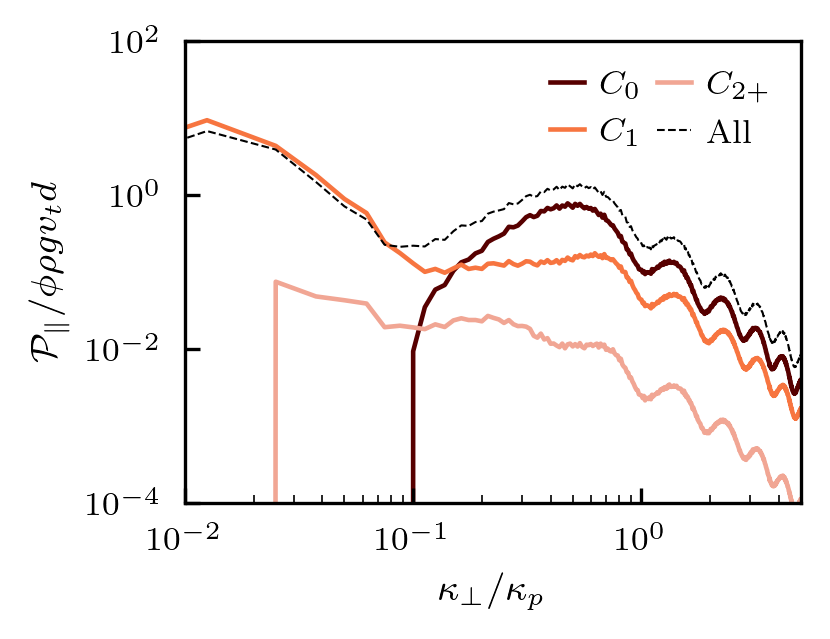}
        \subcaptionlistentry{}
        \put(03,73){\captiontext*{}}
    \end{overpic}%
    \label{fig:cfor100}
  \end{subcaptiongroup}
  \captionsetup{subrefformat=parens}
  \caption{Scale-by-scale variation of the particle power flux along the vertical direction for \subref{fig:cfor10} $\rho=10$ and \subref{fig:cfor100} $100$. The total across all particles $\mathcal{P}_\ver$ is shown using dashed ({\protect\mydashedline}) lines. Continuous lines show the contributions of each cluster $C_i$ according to \cref{eq:forcingdecomp}; $C_{2+}$ indicates the sum of the contributions of all clusters smaller than $C_1$.}
  \label{fig:fig3}
\end{figure*}

\vspace{-\baselineskip}%
\begin{equation}
  \mathcal{P}_{\ver}(\bm{\kappa})=\frac{1}{2}\left(\hat{\bm{u}}^{*}_{\ver}\kern-2pt\cdot\hat{\bm{f}}_{p,\ver}+\hat{\bm{u}}_\ver\kern-2pt\cdot\hat{\bm{f}}_{p,\ver}^{*}\right).
  \label{eq:forcingterm}
\end{equation}
\Cref{fig:fig3} displays the horizontal spectrum of the vertical power flux $\mathcal{P}_\ver(\krad) = \int
\mathcal{P}_\ver\left(\bm{\kappa}\right) \delta_{\rad} d\bm{\kappa}$ normalized by the particle-scale power input. 
The data reveal that the dispersed phase actively injects energy across all resolved scales. To isolate the hydrodynamic footprint of the distinct clusters, we decompose the total momentum forcing field according to our topological Voronoi classification. Since particle stresses have compact support, we can partition the global particle forcing term across the $N$ clusters as
\par\vspace{-\baselineskip}%
\begin{align}
  \hat{\bm{f}}_p (\bm{\kappa})= \mathcal{F}\left(\sum_{i=1}^\ndp \bm{F}_p^{i}\right)
                 = \sum_{i=0}^N\hat{\bm{f}}_p^{C_i}(\bm{\kappa}),
  \label{eq:clusterdecomp}
\end{align}
where $\hat{\bm{f}}_p^{C_i}$ is the spectral forcing purely associated with particles belonging to the cluster $C_i$. Substituting this decomposition into equation \eqref{eq:forcingterm}, we express the global power flux as a linear superposition of contributions from each cluster: 
\par\vspace{-\baselineskip}%
\begin{align}
  \mathcal{P}_{\ver}(\bm{\kappa})= \sum_{i=0}^N \mathcal{P}^{C_i}_{\ver}(\bm{\kappa}).
  \label{eq:forcingdecomp}
\end{align}
The outcome of this spectral partitioning is presented in \cref{fig:fig3}, which contrasts the isolated contributions of the unclustered background $C_0$, the primary cluster $C_1$, and the sum of all smaller satellite clusters $C_{2+}$. In these dense conditions, this decomposition reveals that large-scale energy injection is driven almost entirely by the $C_1$ object: the total large-scale power flux is virtually indistinguishable from the isolated $C_1$ contribution, despite it accounting for a relative minority of the particles. This clearly identifies this structure as the soruce of CIT. In contrast, the power input of the unclustered phase $C_0$ dominates the injection of energy at small scales, peaking just above the particle diameter in accordance with isolated wake dynamics. Satellite clusters $C_{2+}$ instead contribute negligibly to the overall power input. Note that, the contributions of $C_0$ and $C_{2+}$ drop below zero at large scales, indicating that the unclustered particles and small clusters actually absorb energy, further solidifying the $C_1$ cluster as the sole large-scale energy source.
Crucially, the $C_1$ forcing signature at large scales and the overall contribution at small scales both exhibit a power-law scaling regime. Their spectral separation clearly mirrors the fluid kinetic energy spectra ($E_\parallel$). 

The study of the complete energy balance reveals that the transition between these two regimes is delimited by a switch in the energy absorption mechanism. 
For brevity, we focus solely on the distribution of the total energy, $E = E_\ver + E_\rad$. Under statistically steady conditions, the vertically integrated balance is expressed as ${\mathcal{N}(\krad)+\mathcal{D}(\krad)+\mathcal{P}(\krad)=0}$, with the spectral terms defined as

\par
\begin{figure*}[t]
  \centering
  \hrule
  \vspace{1pt}
  \begin{subcaptiongroup}
    \begin{overpic}[width=\linewidth]{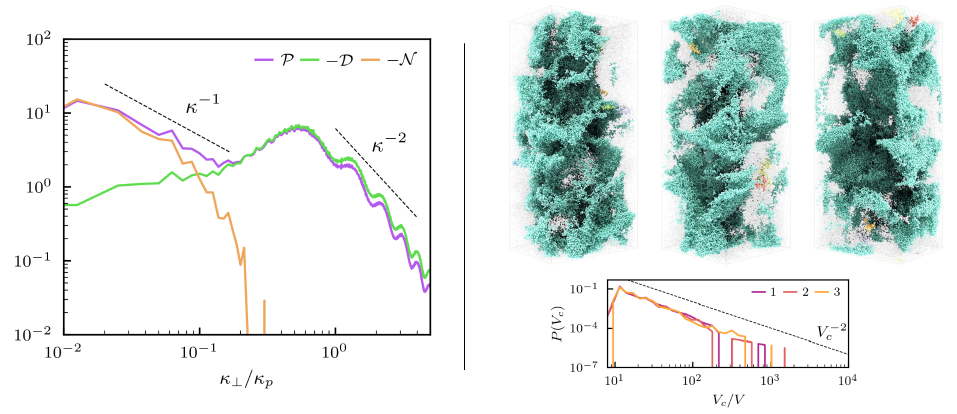}
      \subcaptionlistentry{}
      \label{fig:embud10}
      \put(-01,40){\captiontext*{}}

      \subcaptionlistentry{}
      \label{fig:suppclust}
      \put(48,40){\captiontext*{}}

      \subcaptionlistentry{}
      \label{fig:clustpdfhist}
      \put(48,13){\captiontext*{}}

    \end{overpic}
  \end{subcaptiongroup}
  \vspace{-10pt}
  \hrule
  \captionsetup{subrefformat=parens}
  \caption{Scale-by-scale energy budget in the horizontal plane for $\phi=0.10$, $\rhoratio=10$ (\,\protect\dotA\,). As in \cref{fig:fig3}, all budget terms are normalized by $\phi\rhoratio g \vt d$. \subref{fig:suppclust}~Cluster visualizations for $\phi=0.10$, $\rhoratio = 100$ (\,\protect\dotD\,). Snapshots are spaced approximately $150\ddp/\vt$. Particles belonging to clusters $C_1$--$C_6$ are colored according to the cluster they belong to, whereas particles that do not belong to any cluster $(C_0)$ are shown in semi-transparency. 
  The rightmost sample is the same as that of \cref{fig:cluster}. \subref{fig:clustpdfhist}~Instantaneous cluster distribution for the cases shown in \subref{fig:suppclust}; numbering (1--3) is from left to right. Due to the dominance of $C_1$, the each distribution shows singular peaks at high $V_\cluster/V_\voro$. }
  \label{fig:app}
\end{figure*}

\vspace{-\baselineskip}%
\begin{align*}
  &E(\krad) = \int \hat{\bm{u}}\cdot\hat{\bm{u}}^{*} \delta_{\rad} \text{d}^3\bm{\kappa}, & 
  &\mathcal{N}(\krad) = -i \int \bm{\kappa}\cdot\left(\widehat{\bm{u}\otimes\bm{u}}\hat{\bm{u}}^{*} + \widehat{\bm{u}\otimes\bm{u}}^{*}\hat{\bm{u}}\right) \delta_{\rad} \text{d}^3\bm{\kappa},\\
  &\mathcal{D}(\krad) = -2 \nu \krad^2 E(\krad), & 
  &\mathcal{P}(\krad) = \frac{1}{\rhof} \int \left(\hat{\bm{f}}_p\cdot\hat{\bm{u}}^{*} + \hat{\bm{f}}_p^{*}\cdot\hat{\bm{u}}\right) \delta_{\rad}\text{d}^3\bm{\kappa};
  \label{eq:terms}
\end{align*}
 a full derivation from the Fourier-transformed Navier-Stokes equations is reported by \cite{pope2000}. 
By definition, negative values in the spectral budget indicate kinetic energy extraction from a given scale, while positive values denote energy injection. The resulting balance is shown in \cref{fig:embud10} for the $\phi=0.10$, $\rhoratio=10$ case. We observe that the nonlinear term $\mathcal{N}$ acts as an energy sink ($\mathcal{N} < 0$), balancing the injected energy from the particles $\mathcal{P}$ and dominating over the viscous dissipation $\mathcal{D}$, precisely at the large scales characterized by the $C_1$ forcing signature. 
Conversely, at the particle scale and below, nonlinear transfers are negligible, and the energy injected by the particles is directly balanced by local viscous dissipation.

A simple scaling analysis connects the spectral budget to the energy spectra. As shown in \cref{fig:ekx}, the spectrum exhibits two distinct regions of approximately self-similar scaling, where $E(\krad) \sim \krad^{\alpha}$: a high-wavenumber regime ($\krad \gtrsim \kappa_p$) dominated by particle wakes, and a large-scale regime ($\krad < \kappa_p$) characterized by CIT. The particle energy injection exhibits two corresponding scaling regions where $\mathcal{P}(\krad) \sim \krad^{\beta}$ (\cref{fig:fig3,fig:embud10}), so that:
\begin{enumerate}[label=\roman*)]
  \item at high wavenumbers ($\krad \gtrsim \kappa_p$), where nonlinear transfers are negligible, the budget reduces to a local balance between injection and dissipation, $\mathcal{P} \approx -\mathcal{D}$. Since $\mathcal{D} \propto \krad^2 E(\krad)$, this balance requires $\krad^\beta \propto \krad^{2+\alpha}$, yielding $\beta = 2 + \alpha$. Using the measured energy spectrum exponent $\alpha \approx -4$ (\cref{fig:ekx}), this relation predicts an injection scaling of $\beta \approx -2$, consistent with the measurements in \cref{fig:embud10}.
  \item At small wavenumbers ($\krad < \kappa_p$), the budget is governed by a balance between energy injection and nonlinear flux, $\mathcal{P} \approx -\mathcal{N}$. By dimensional arguments, the nonlinear spectral transfer must scale as $\mathcal{N}(\krad) \sim \krad^{3/2} E(\krad)^{3/2}$. Equating this to the injection spectrum $\mathcal{P}(\krad) \sim \krad^{\beta}$ and substituting the observed energy scaling $E(\krad) \sim \krad^{\alpha}$, we obtain $\krad^{\beta} \propto \krad^{3(\alpha+1)/2}$ and the theoretical prediction $\beta = \frac{3}{2}(\alpha+1)$. As shown in \cref{fig:ekx}, the measured energy spectrum exhibits a scaling of $\alpha \approx -1.7$ and consequently, the model predicts an injection scaling of $\beta \approx -1$, confirmed by the budget measurements in \cref{fig:embud10}.
\end{enumerate}
While shown only for the $\rhoratio=10$ case, these scaling relations hold for the $\rhoratio=100$ case as well.

These results provide direct evidence of the self-similar energy injection governed by the fractal geometry of the cluster. The spectral wavenumber interval of this self-similar forcing closely matches the spatial scaling range of the cluster's fractal geometry (\cref{fig:fdim}), establishing a direct, mechanistic link between the multiscale topology of the primary cluster and its corresponding fluid excitation. While this topological shape is invariant, the exponent of the forcing steepens at higher $\rho$ (\cref{fig:fig3}), confirming that particle inertia dictates the intensity of the energy transfer.

\vspace{-10pt}
\section{Conclusions}
In summary, using state-of-the-art PR-DNS in large domains, we have provided the first fully resolved demonstration of cluster-induced turbulence (CIT) emerging spontaneously from the sedimentation of inertial particles, in the absence of externally imposed background turbulence. We measure the emergence of a broadband, quasi-isotropic fluid kinetic energy distribution. At high volume fractions ($\phi=0.10$) the spectra evidence both a low wavenumber region ($\krad<\kappa_p$) of strong self-similar scaling, and the distinctly separate, previously  known particle-scale fluctuations ($\krad\gtrsim\kappa_p$). The spectral energy budget reveals that the transition between the two regimes corresponds to a switch in the energy absorption mechanism between the nonlinear and dissipation terms. 
A topological analysis of the disperse phase shows a steep, self-similar distribution of cluster volumes ($V_\cluster^{-2.5}$), and a strict structural hierarchy, overwhelmingly dominated by a single, highly connected, domain-spanning, macroscopic fractal cluster.  
A spectral decomposition of the momentum coupling term of the energy budget shows a cluster-scale power law injection that mirrors the geometric self-similarity of the disperse phase. 

The self-similar forcing of turbulent fluctuations has previously been investigated by fractal grids and analyzed via renormalization-group theory \citep{biferale-lanotte-ea2004,mazzi-vassilicos2004}. However, standard theoretical scaling relations connecting static forcing exponents to velocity fluctuations do not fully capture the dynamics observed here. This discrepancy highlights the fundamentally novel physics of CIT: unlike static fractal grids, these clusters exist in a highly dynamic, two-way coupled relationship with the continuous phase, constantly reshaping themselves in response to the very turbulence they generate; we show examples of the emerging structures in \cref{fig:suppclust,fig:clustpdfhist}. 

They also highlight the need to account for the multiscale structure and dynamics of the dispersed phase when developing predictive models of multiphase flows. 
The present investigation covers a computationally tractable region of the ($\phi$,\,$\rho$,\,$\Ret$) parameter space, chosen to enable the spontaneous development of CIT while resolving the relevant particle-scale dynamics. Extending this analysis to higher mass loadings, broader parameter ranges and larger domains will establish how robust the observed scaling laws are and how the forcing spectrum depends on the flow conditions. More broadly, our results identify dynamically evolving, self-similar particle clusters as an intrinsic mechanism for energy injection in multiphase turbulence, opening a route toward predictive descriptions that connect particle organization, interphase momentum transfer and turbulent dynamics.

\input{appendix.tex}

\begin{bmhead}[Acknowledgments]
{\footnotesize The research was supported by the Okinawa Institute of Science and Technology Graduate University (OIST) with subsidy funding to M.E.R. from the Cabinet Office, Government of Japan. M.E.R. also acknowledges funding from the Japan Society for the Promotion of Science (JSPS), grant 24K17210 and 24K00810. The authors acknowledge the computer time provided by the Scientific Computing \& Data Analysis section of the Core Facilities at OIST, and by HPCI, under the Research Project grants hp250021 and hp260019. S.T. benefited from valuable discussions with P. Garg and L. Fossà while preparing this work.} 
\end{bmhead}

\begin{bmhead}[Declaration of Interests]
{\footnotesize The authors report no conflict of interest.}
\end{bmhead}

\vspace{-10pt}
\bibliographystyle{jfm}
\bibliography{biblio}

\end{document}